\documentstyle[prl,aps,multicol,epsf]{revtex}
\def \I{{\rm i}}
\def \I{{\rm i}}
\def \E{{\rm e}}

\def \b#1{{${\hat b}_#1$}}

\begin{document}


\title{Pulse-mode quantum projection synthesis: Effects of mode mismatch on optical state truncation and preparation}

\author{\c{S}ahin Kaya \"Ozdemir,$^{(a)}$ Adam Miranowicz,$^{(a,b)}$ Masato Koashi,$^{(a)}$ and Nobuyuki
Imoto$^{(a,c,d)}$}
\address{
$(a)$ CREST Research Team for Interacting Carrier Electronics,
\\ The Graduate University for Advanced
Studies (SOKEN-DAI), Hayama, Kanagawa 240-0193, Japan\\
$(b)$ Nonlinear Optics Division, Institute of Physics, Adam
Mickiewicz
University, 61-614 Pozna\'n, Poland\\
$(c)$ NTT Basic Research Laboratories, 3-1 Morinosato Wakamiya, Atsugi, Kanagawa 243-0198, Japan\\
$(d)$ Department of Applied Physics, University of Tokyo, 7-3-1 Hongo, Bunkyo-ku, Tokyo 113-8654, Japan}

\date{\today}
\pagestyle{plain} \pagenumbering{arabic} \maketitle
 \widetext
\begin{abstract}
Quantum projection synthesis can be used for phase-probability-distribution measurement, optical-state
truncation and preparation. The method relies on interfering optical lights, which is a major challenge in
experiments performed by pulsed light sources. In the pulsed regime, the time frequency overlap of the
interfering lights plays a crucial role on the efficiency of the method when they have different mode
structures. In this paper, the pulsed mode projection synthesis is developed, the mode structure of
interfering lights are characterized and the effect of this overlap (or mode match) on the fidelity of
optical-state truncation and preparation is investigated. By introducing the positive-operator-valued measure
(POVM) for the detection events in the scheme, the effect of mode mismatch between the photon-counting
detectors and the incident lights are also presented.
\end{abstract}
 \widetext
\begin{multicols}{2}
\section{Introduction}
The accurate preparation of quantum states is a crucial task for reliable quantum computation and
quantum-information processing. Several schemes have been proposed for the generation of arbitrary states and
their superpositions. One of the most developed systems of state preparation relies on \textit{conditional
measurement}, which brings one of the subsytems of an entangled system to a predetermined state by a
measurement on the other subsystem. In these systems, entanglement of the two subsystems is achieved through
linear or nonlinear interactions \cite{state,Barnett98b,Dak99,Dar00}.

The \textit{projection-synthesis} approach, which has been originally proposed to measure the optical phase
probability distribution by Barnett et al. \cite{Barnett96}, exploits the mixing of two states (one to be
measured and the other as reference state) at a beam splitter and a measurement at the output states of the
beam splitter \cite{Moussa97,Barnett98a}. This approach, despite of its simplicity, is very flexible to be
used for different applications among which we can count the optical state truncation
\cite{Barnett98b,Sahin01,Kon00,Villas00}, preparation of superposition and phase states
\cite{Paris01,Sahin02,Villas01} and the teleportation of superposition states \cite{Villas99}. The scheme,
which is shown in Fig.\ref{Fig1}, exploits projection synthesis and is often referred to as \textit{Quantum-
Scissors Device} (QSD) in the applications of optical-state truncation and preparation. It relies on linear
optical elements (two beam splitters, BS), a single-photon state, a coherent state and two photon-counting
detectors. In the first beam splitter (BS1), single-photon state is mixed with vacuum and an entangled state
of one-photon state and vacuum is formed at the output ports of BS1. The state at one of the output ports of
BS1 is sent to the second beam splitter (BS2), where it is mixed with the input coherent light to be
truncated. The photon-counting detectors placed at the output ports of BS2 count the number of photons
incident on them after the action of BS2. The state at the other output port of BS1 is projected on a
specific state among many others according to the number of photons counted at the detectors. In the special
case of one-photon detection by one of the detectors and none by the other, the output is projected onto a
superposition of vacuum and one-photon state which carries the relative phase and amplitude information of
the vacuum and one-photon components of the input coherent state.

As is the case for any scheme, where the interference of differently processed light pulses takes place, the
characterization of the optical modes of these lights and their effects on the outcome of the experiments is
a major challenge for the quantum-scissors device, too. Although it has been shown that the scheme is
realizable with the current level of quantum optics technology \cite{Sahin01}, the studies so far have not
considered the problem of mode matching. In this paper, we investigate projection synthesis for
quantum-scissors device using the pulse mode formalism and study the effect of mode-mismatch problem on state
truncation and preparation.

For the evaluation of the quality of the process, fidelity of the generated state to the desired one is used.
Fidelity is a commonly used measure of how close the two states are and is given by
\begin{equation}\label{N01}
F={\rm Tr}[\hat{\rho}_{{\rm out}}|\phi_{{\rm desired}}\rangle\langle\phi_{{\rm desired}}|]~,~
\end{equation}
where $\hat{\rho}_{{\rm out}}$ and $|\phi_{{\rm desired}}\rangle$ are the prepared and desired states,
respectively. When the prepared state is exactly the desired state then $F=1$, when these two states are
orthogonal $F=0$. In practice, the value of fidelity will lie between $0$ and $1$, and its value will be a
sign of the quality of the process. In general, the quality of the prepared state strongly depends on the
details of the mixing (interference) process and the conditional measurement. When these two main phenomena
are prone to errors, the generated state may considerably differ from the desired one.

The paper is organized as follows: In Sec. II, pulse-mode formalism is introduced and the calculation of mode
mismatch is explicitly shown. The effects of mode mismatch between the interfering lights and the
photodetectors are studied in detail in Sec. III,  and analytical expressions, which show the mismatch
dependence of fidelity of state truncation by projection synthesis, are given. Then in Sec. IV, the results
of the findings are discussed for preparation of arbitrary superposition of vacuum and one-photon states. A
discussion of some practical issues and the characterization of mode structures of fields in a practical
scheme are addressed in Sec. V. And finally, Sec. VI includes a brief summary and conclusion of this study.

\section{Theory of Mode Mismatch Using Pulse Mode Formalism} 
\begin{figure*}[h]
\vspace*{-25mm} \hspace*{-20mm} \epsfxsize=12cm \epsfbox{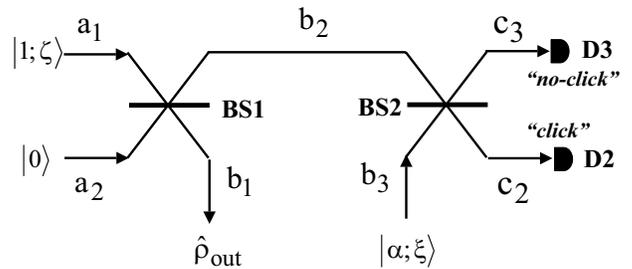}\vspace{-24mm} \caption{Schematic
configuration of the quantum-scissors device (QSD). BS1, BS2: beam splitters; D2, D3: photon-counting
detectors; $|\alpha\rangle, |0\rangle, |1\rangle$: coherent, vacuum and single-photon states, respectively,
$\zeta$ and $\xi$ denote the mode functions of the corresponding input fields and $\rho_{out}$ is the
truncated output state. Desired output state is obtained by one-photon detection at D2 which is denoted as
``click" and no photon detection (``no-click") at D3. \label{Fig1}}
\end{figure*}

In the QSD scheme shown in Fig.\ref{Fig1}, interference
of vacuum and single-photon states at BS1 (50:50) and that of the entangled state of mode \b2 mode and the
coherent state at BS2 (50:50) are the fundamental optical processes. The scheme is usually analyzed in the
single mode description in which a pair of annihilation and creation operators for each beam splitter is
used. In that picture, the spatio-temporal characteristics of the states input to the beam splitters are
assumed to be matched perfectly at the beam splitters and detectors. However, in practice, these interfering
lights are prepared independently and thus may have different modes. Moreover, mode definitions of the states
at the output of BS2 and that of the measuring apparatus (photon counting detectors) may be different. In
these cases, the detection of the correct photon numbers does not mean the correct conditioning (projection)
of the desired output state. In practical experiments, high level of attention must be given to match the
modes of the input states and the detectors as much as possible for a successful state preparation. A good
mode matching shows itself as high visibility and can be a major challenge in experiments.

In this section, we will introduce the pulse-mode formalism and present general expressions to calculate the
overlap of two number states with different modes. It is assumed that the bandwidth of the light pulses are
sufficiently small and the variation of the beam-splitter transmission and reflection parameters within the
pulse bandwidths can be neglected. In this case, these parameters become independent of pulse shape and
solely reflect
\noindent the effect of beam splitters; beam splitting will not affect the mode structure of the input
light pulses.

Following Refs. \cite{Blow,Hussain,Imoto,Santos}, we define creation and annihilation operators of light
pulses in terms of the operators of the monochromatic modes that form them. Then the creation operator for a
pulse whose mode profile is described by $\xi$ can be written as
\begin{eqnarray}\label{N10}
\hat{a}^{\dagger}(\xi)=\int d\omega \xi(\omega)\hat{a}^{\dagger}(\omega)~,
\end{eqnarray}
\noindent where $\xi(\omega)$ is normalized as
\begin{equation}\label{N11}
\|\xi\|^{2}\equiv\int d\omega|\xi(\omega)|^{2}=1~.
\end{equation}
\noindent Using the continuous-mode bosonic commutators
\begin{equation}\label{N12}
[\hat{a}(\omega),\hat{a}^{\dagger}(\omega^{'})]=\delta(\omega-\omega^{'})~,
\end{equation}
\noindent we can obtain the following commutator
\begin{equation}\label{N13}
[\hat{a}(\xi),\hat{a}^{\dagger}(\xi)]=1~.
\end{equation}
\noindent

In order to analyze the effects of mode mismatch, we have to look at the relation between the mode
descriptions of creation operators. The commutator between two operators with different mode descriptions of
$\xi(\omega)$ and $\zeta(\omega)$ can be calculated as follows \cite{Imoto}
\begin{eqnarray}\label{N14}
  [\hat{a}(\xi),\hat{a}^{\dagger}(\zeta)]&=&\left[\int{d\omega\xi^{\ast}(\omega) \hat{a}(\omega)} ,
  \int{d\omega'\zeta(\omega') \hat{a}^{\dag}(\omega')} \right]\nonumber
  \\&=&\int{d\omega}\int{d\omega'\xi^{\ast}(\omega)\zeta(\omega')[\hat{a}(\omega),\hat{a}^{\dagger}(\omega')]}\nonumber
  \\&=&\int{d\omega}\xi^{\ast}(\omega)\zeta(\omega)=(\xi,\zeta)~,
\end{eqnarray}
\noindent which means that the commutator between operators of any two modes corresponds to the overlap of
these two modes.

The operators defined above can be used to construct number and coherent states with a given mode description
simply by replacing the usual discrete bosonic operators with the pulse-mode operators of the given mode
description. In the following, $|\sqcup;\xi\rangle$ represents a Fock state when $\sqcup$ is a number or
written in roman, and a coherent state when $\sqcup$ is written in Greek alphabet. $\xi$ denotes the
mode-profile of the corresponding state. Then a number state of mode $\xi$ can be written as
\begin{equation}\label{N19}
|n;\xi\rangle=\frac{1}{\sqrt{n!}}[\hat{a}^{\dagger}(\xi)]^{n}|0\rangle~,
\end{equation}
\noindent and a coherent state as
\begin{equation}\label{N20}
|\alpha;\xi\rangle=\exp[\alpha \hat{a}^{\dagger}(\xi)-\alpha^{*}\hat{a}(\xi)]|0\rangle~
\end{equation} \noindent with $\hat{a}(\omega)|\alpha;\xi\rangle=\alpha \xi(\omega)|\alpha;\xi\rangle$. Here,
we define the mode profile function as
\begin{equation}\label{N6789}
g(\omega,\omega')={\rm Tr}[\hat{\rho}\hat{a}^{\dagger}(\omega)\hat{a}(\omega')]
\end{equation}\noindent which gives $g(\omega,\omega')=\zeta(\omega,\omega')=\zeta(\omega)\zeta^{*}(\omega')$ for a
one-photon state $\hat{\rho}=|1;\zeta\rangle\langle1;\zeta|$, and
$g(\omega,\omega')=\xi(\omega,\omega')=|\alpha|^{2}\xi(\omega)\xi^{*}(\omega')$ for a single mode coherent
state $\hat{\rho}=|\alpha;\xi\rangle\langle\alpha;\xi|$ with $\int d\omega\zeta(\omega,\omega)=1$ and $\int
d\omega\xi(\omega,\omega)=|\alpha|^2$.

The overlap $\langle n;\xi|n;\zeta\rangle$ of two pure number states $|n;\zeta\rangle$ and $|n;\xi\rangle$
can be found by successive applications of Eqs. (\ref{N10})-(\ref{N14}):
\begin{eqnarray}\label{aaa2}
\langle n;\xi|n;\zeta\rangle&=&\prod_{k=1}^{n}\int d\omega _{k}\xi ^{\ast }\left( \omega _{k}\right) \zeta
\left( \omega _{k}\right) \nonumber \\
&=&\prod_{k=1}^{n}(\xi ,\zeta )=(\xi ,\zeta )^{n}
\end{eqnarray} \noindent from which the overlap of two pure single photon states can be found as
$\langle 1;\xi|1;\zeta\rangle=(\xi ,\zeta )$.

The overlap of a pure one-photon state $|1;\xi\rangle$ and a mixed one $\hat{\rho}_{1}$, which is defined as
\begin{eqnarray}\label{bbb}
&&\hat{\rho}_{1}=\sum_{j} p_{j}~|1;\zeta_{j}\rangle\langle 1;\zeta_{j}|
\end{eqnarray}
with $\sum_{j} p_{j}=1$, is found using ${\rm Tr}[\hat{\rho}_{1}|1;\xi\rangle \langle 1;\xi|]$ as
\begin{eqnarray}\label{bbb6}
&&\langle 1;\xi|\hat{\rho}_{1}|1;\xi\rangle =\int \int d\omega d\omega'\sum_{j}p_{j}\zeta_{j}
^{\ast}(\omega')\xi(\omega')\zeta_{j}(\omega)\xi^{\ast}
(\omega)\nonumber \\
&&~~~~~~~~~~~~~~~~~~~~~=\int \int d\omega d\omega'\zeta(\omega,\omega')\xi(\omega,\omega')\nonumber \\
&&~~~~~~~~~~~~~~~~~~~~~=\sum_{j}~p_{j}~|(\xi ,\zeta_{j} )|^{2}
\end{eqnarray} \noindent where $\zeta(\omega,\omega')=\sum_{j}p_{j}\zeta_{j}(\omega)\zeta_{j} ^{\ast }(\omega')$, and
$\xi(\omega,\omega')=\xi(\omega')\xi^{\ast }(\omega)$ are found using Eq. (\ref{N6789}).

The overlap between a mixed number state $\hat{\rho}_{1}$ as given in Eq. (\ref{bbb}) and a pure coherent
state $|\alpha;\xi\rangle$ can be found first writing the coherent state in the photon number basis and then
applying the above procedure. This will result in
\begin{eqnarray}\label{kkk2}
&&\langle \alpha;\xi|\hat{\rho}_{1}|\alpha;\xi\rangle=\sum_{j}~p_{j}\exp[-|\alpha|^{2}]\times|(\zeta_{j},\xi)\alpha|^{2} \nonumber \\
&&~~~~~~=(\Lambda)^{-1} \int \int d\omega
d\omega'\underbrace{|\alpha|^{2}\xi(\omega')\xi^{\ast}(\omega)}_{\xi(\omega,\omega')}\nonumber\\
&&~~~~~~~~~~~~~~~~~~~\times \underbrace{\sum_{j}~p_{j}\zeta_{j}(\omega)\zeta_{j}
^{\ast}(\omega')}_{\zeta(\omega,\omega')}\nonumber\\
&&~~~~~~=\frac{\int \int d\omega d\omega'{\xi(\omega,\omega')}{\zeta(\omega,\omega')}}{\int
d\omega{\xi(\omega,\omega)} \int d\omega{\zeta(\omega,\omega)}}
\end{eqnarray}\noindent with $\Lambda=\int d\omega \xi(\omega,\omega)\int d\omega \zeta
(\omega,\omega)=|\alpha|^2$. Here we define Eq. (\ref{kkk2}) as the mode-match parameter $|\gamma_{0}|^2$
which satisfies $|\gamma_{0}|^{2}+|\gamma_{1}|^{2}=1$ with $|\gamma_{1}|^{2}$ representing the mode-mismatch
parameter.

In the following sections, we will use the formalism developed in this section to study the QSD scheme where
the input fields are a mixed single photon state and a pure coherent state.

\section{Analysis of Mode Mismatch in QSD Scheme for State Truncation}

In an optical-state truncation experiment using the QSD scheme, an input coherent light with an unknown
quantum state (intensity and phase) is truncated up to its one-photon state generating, at a remote port, a
superposition of its vacuum and one-photon state preserving the relative phase and intensity between these
components of the input coherent light.

In this section, we study the state truncation in the pulsed regime and investigate the effect of mode
mismatch between the interfering lights on the fidelity of the truncation process. We also consider the case
where the mode structures of the photon-counting detectors are different from the mode structures of the
light pulses incident on them. We assume that the input coherent light, with unknown quantum state, is at
$\hat{b}_{3}$ input of BS2 and has a mode profile described by $\xi$. The $\hat{a}_{1}$ input port of BS1 is
fed with a single-photon state whose mode profile is given by $\zeta$. First, we consider the case where the
one-photon input to the device is prepared in a mixed state and find the general expressions for the output
density operator and fidelity of the process. After the presentation of the general formulas, we will analyze
the process, in details, for a single photon prepared in pure state and present the results of this study. In
the evaluation of the efficiency of the truncation process in this section, we will impose the condition that
the output state should be a superposition of vacuum and one-photon states $N(|0\rangle+\alpha|1\rangle)$ in
the same mode of the input coherent light $|\alpha\rangle$.

With the single-photon prepared in a mixed state
\begin{equation}\label{h1}
\hat{\rho}_{a_{1}}=\sum_{j} p_{j}~|1;\zeta_{j}\rangle _{a_{1} a_{1}}\langle 1;\zeta_{j}|
\end{equation}\noindent with $\sum_{j} p_{j}=1$, and the coherent state as
$|\alpha;\xi\rangle_{b_{3}}$, the overall input to the QSD scheme
becomes
\begin{eqnarray}\label{h2}
\hat{\rho}_{\rm in}=\sum_{j}p_{j}|1;\zeta_{j}\rangle _{a_{1} a_{1}}\langle 1;\zeta_{j}|\otimes|0\rangle
_{a_{2} a_{2}}\langle0|\otimes|\alpha;\xi\rangle_{b_{3}b_{3}}\langle \alpha;\xi|.\nonumber
\end{eqnarray}\noindent
Then the state state just before the photon counting can be written as $
\hat{\rho}_{(b_{1},c_{2},c_{3})}=\hat{U}^{\dagger}_{2}\hat{U}^{\dagger}_{1}\hat{\rho}_{\rm
in}\hat{U}_{1}\hat{U}_{2}$ where the actions of the beam splitters BS1 and BS2 are represented by unitary
operators $\hat{U}_{1}$ and $\hat{U}_{2}$, respectively \cite{Sahin01,Sahin02,Campos}.

The probability of detecting a ``click" at D2 and ``no-click" at D3 is given by the trace over the three
modes
\begin{equation}\label{N22}
{\rm P}_{10}={\rm Tr}_{(b_{1},c_{2},c_{3})}[~\hat{\rho}_{(b_{1},c_{2},c_{3})}~{\rm \Pi}_{1}^{c_{2}} {\rm
\Pi}_{0}^{c_{3}}~]
\end{equation}
with ${\rm \Pi}_{1}^{c_{2}}$ and ${\rm \Pi}_{0}^{c_{3}}$ being the elements of positive-operator-valued
measures (POVMs). In general for a detector with a quantum efficiency of $\eta$, the POVM can be written as
\begin{eqnarray} {\rm \Pi}_{n}= \sum_{m=n}^\infty \eta^{n}(1-\eta)^{m-n}C_{n}^{m}|m\rangle
\langle m|~, \label{N38}
\end{eqnarray}
\noindent  where $n$ and $m$ are the number of detected and incident photons, respectively \cite{BarnettOpt}.
$C_{n}^{m}$ represent the binomial coefficients, and $\sum_{0}^{\infty}{\rm \Pi}_{n}=1$. For the sake of
simplicity, we assume zero mean dark count ($\nu=0$) in this study.

Then the output state at $\hat{b}_{1}$ which is conditioned on this detection is found by a partial trace
\begin{eqnarray}\label{N23}
\hat{\rho}_{\rm out}=\frac{1}{{\rm P}_{10}}{\rm Tr}_{(c_{2},c_{3})}[~\hat{\rho}_{(b_{1},c_{2},c_{3})}~{\rm
\Pi}_{1}^{c_{2}} {\rm \Pi}_{0}^{c_{3}}~].
\end{eqnarray} \noindent Then the density operator is written as
\begin{eqnarray}\label{N23abcd}
\hat{\rho}_{\rm out}&=&\frac{1}{{\rm
P}_{10}}{\sum_{j}p_{j}\big{[}}~d_{00}^{(j)}|0\rangle_{b_{1} b_{1}}\langle0|+d_{01}^{(j)}|0\rangle_{b_{1} b_{1}}\langle1;\zeta_{j}|\nonumber \\
&~&~~~~~~~+d_{10}^{(j)}|1;\zeta_{j}\rangle_{b_{1}
b_{1}}\langle0|+d_{11}^{(j)}|1;\zeta_{j}\rangle_{b_{1}b_{1}}\langle1;\zeta_{j}|~{\big{]}}
\end{eqnarray} \noindent with the following elements
\begin{eqnarray}\label{N24}
d_{00}^{(j)}&=&\frac{1}{4}~_{c_{2}}\langle\lambda;\xi|\hat{c}_{2}(\zeta_{j}){\rm
\Pi}_{1}^{c_{2}}\hat{c}^{\dag}_{2}(\zeta_{j})|\lambda;\xi\rangle_{c_{2}c_{3}}\langle\delta;\xi|{\rm
\Pi}_{0}^{c_{3}}|\delta;\xi\rangle_{c_{3}}\nonumber \\
&+&\frac{1}{4}~_{c_{2}}\langle\lambda;\xi|{\rm
\Pi}_{1}^{c_{2}}|\lambda;\xi\rangle_{c_{2}c_{3}}\langle\delta;\xi|\hat{c}_{3}(\zeta_{j}){\rm
\Pi}_{0}^{c_{3}}\hat{c}^{\dag}_{3}(\zeta_{j})|\delta;\xi\rangle_{c_{3}}\nonumber \\
&+&\frac{i}{4}~_{c_{2}}\langle\lambda;\xi|\hat{c}_{2}(\zeta_{j}){\rm
\Pi}_{1}^{c_{2}}|\lambda;\xi\rangle_{c_{2}c_{3}}\langle\delta;\xi|{\rm
\Pi}_{0}^{c_{3}}\hat{c}^{\dag}_{3}(\zeta_{j})|\delta;\xi\rangle_{c_{3}}\nonumber \\
&-&\frac{i}{4}~_{c_{2}}\langle\lambda;\xi|{\rm
\Pi}_{1}^{c_{2}}\hat{c}^{\dag}_{2}(\zeta_{j})|\lambda;\xi\rangle_{c_{2}c_{3}}\langle\delta;\xi|\hat{c}_{3}(\zeta_{j}){\rm
\Pi}_{0}^{c_{3}}|\delta;\xi\rangle_{c_{3}}\nonumber \\
d_{01}^{(j)}&=&\frac{-1}{2\sqrt{2}}~_{\hat{c}_{2}}\langle\lambda;\xi|{\rm
\Pi}_{1}^{c_{2}}|\lambda;\xi\rangle_{c_{2}c_{3}}\langle\delta;\xi|{\rm
\Pi}_{0}^{c_{3}}\hat{c}^{\dag}_{3}(\zeta_{j})|\delta;\xi\rangle_{c_{3}}\nonumber \\
&+&\frac{i}{2\sqrt{2}}~_{c_{2}}\langle\lambda;\xi|{\rm
\Pi}_{1}^{c_{2}}\hat{c}^{\dag}_{2}(\zeta_{j})|\lambda;\xi\rangle_{c_{2}c_{3}}\langle\delta;\xi|{\rm
\Pi}_{0}^{c_{3}}|\delta;\xi\rangle_{c_{3}}\nonumber \\
d_{11}^{(j)}&=&\frac{1}{2}~_{c_{2}}\langle\lambda;\xi|{\rm
\Pi}_{1}^{c_{2}}|\lambda;\xi\rangle_{c_{2}c_{3}}\langle\delta;\xi|{\rm
\Pi}_{0}^{c_{3}}|\delta;\xi\rangle_{c_{3}}
\end{eqnarray} \noindent
and $d_{10}^{(j)}=d_{01}^{(j)*}$ where $\delta=\alpha /\sqrt{2}$ and $\lambda=i\alpha /\sqrt{2}$  are
obtained through the action of the BS2 on $|\alpha;\xi\rangle_{b_{3}}$. The creation operators associated
with the outgoing modes of BS2 are represented by $\hat{c}^{\dag}_{k}$ where $k=2,3$.

Then the fidelity of this output state $\hat{\rho}_{\rm out}$ to the desired truncated state
\begin{equation}\label{N25}
  |\phi_{\rm desired}\rangle=\frac{|0\rangle_{b_{1}}+\alpha|1;\xi\rangle_{b_{1}}}{\sqrt{1+|\alpha|^{2}}}~,
\end{equation} \noindent
which has the same mode profile $\xi$ of the input coherent light, can be calculated, using Eq. (\ref{N01}),
as
\begin{eqnarray}\label{N26}
F=\frac{\sum_{j}p_{j}\left[d^{(j)}_{00}+2{\rm Re}[\alpha \Upsilon_{j}
d^{(j)}_{01}]+d^{(j)}_{11}|\alpha|^{2}|\Upsilon_{j}|^{2}\right]}{(1+|\alpha|^{2})\sum_{j}p_{j}(d^{(j)}_{00}+d^{(j)}_{11})},
\end{eqnarray}\noindent
where $_{b_{1}}\langle1;\zeta_{j}|1;\xi\rangle_{b_{1}}=(\zeta_{j},\xi)=\Upsilon_{j}$  represents the overlap
of the mode of the output single-photon state and that of the desired output state. The effect of the overlap
of the photon-counting detectors and the fields incident on them is contained in the expressions of the
elements of the output density matrix which will be clear in the following subsections.

State truncation using the QSD scheme is based on conditional measurement. Therefore, the correct application
and interpretation  of photodetection process is essential to evaluate this scheme. In the following
subsections, we will present a comparative study of different photon-counting detectors. First, we will use
ideal counters which can resolve the photon number incident on them and then proceed with a realistic
description of photodetection with conventional photon counters.

\subsection{Photon-number-resolving detectors}
This type of detectors can resolve the number of incident photons. In the following, we will first analyze
the scheme for detectors that are matched only to a specific mode and then present the elements of POVM for a
more realistic case where the mode of the incident light cannot be resolved.

\subsubsection{Mode-resolving detectors}

For mode-resolving detectors, the elements of the POVMs can be written as
\begin{eqnarray}\label{Nxx12}
{\rm \Pi}_{0}^{c_{3}}&=& \sum_{m=0}^\infty (1-\eta)^{m}\hat{P}^{c_{3}}_{m}(\varrho)\nonumber \\
{\rm \Pi}_{1}^{c_{2}}&=& \sum_{m=0}^\infty m \eta(1-\eta)^{m-1}\hat{P}^{c_{2}}_{m}(\varrho)
\end{eqnarray}\noindent
\noindent where $\hat{P}^{c_{k}}_{m}(\varrho)$ with $k=2,3$ is the projection onto the eigenspace of
$\hat{c}^{\dag}_{k}(\varrho)\hat{c}_{k}(\varrho)$ with eigenvalue $m$ satisfying the commutators
$[\hat{P}^{c_{k}}_{m}(\varrho),\hat{c}^{\dag}_{k}(\xi)]=0$ and
$[\hat{P}^{c_{k}}_{m}(\varrho),\hat{c}_{k}(\xi)]=0$ if the overlap $(\xi,\varrho)=0$. Here $\varrho$
represents the light mode that can be resolved by the detectors and $\varrho^{\perp}$ represents the
unresolved light mode with $(\varrho,\varrho^{\perp})=0$. Then the light modes in Eq. (\ref{N24}) can be
decomposed into two orthogonal modes as $\zeta_{j}=\chi_{j}\varrho+\Xi_{j}\varrho^{\perp}_{\zeta_{j}}$ and
$\xi=\kappa\varrho+\mu\varrho^{\perp}_{\xi}$ where we define $\chi_{j}=(\varrho,\zeta_{j})$,
$\Xi_{j}=(\varrho^{\perp}_{\zeta_{j}},\zeta_{j})$ with $|\chi_{j}|^{2}+|\Xi_{j}|^2=1$ and $\kappa=(\varrho,
\xi)$, $\mu=(\varrho^{\perp}_{\xi},\xi )$ with $|\kappa|^{2}+|\mu|^2=1$ to represent the overlap (mode match)
of two modes characterized by $\zeta_{j}$ and $\xi$, with the mode $\varrho$ that can be resolved by the
detectors. Consequently, the annihilation and creation operators for a given mode can be decomposed in the
same way resulting in
$\hat{c_{3}}(\xi)=\kappa^{*}~\hat{c_{3}}(\varrho)+\mu^{*}~\hat{c_{3}}(\varrho^{\perp}_{\xi})$. A similar
expression can be obtained for $\hat{c_{2}}(\zeta_{j})$ by using the given relations above. The overlap of
the modes $\varrho^{\perp}_{\zeta_{j}}$ and $\varrho^{\perp}_\xi$ can be found by using the commutators given
in Eqs. (\ref{N13}) and (\ref{N14}) as
\begin{eqnarray}\label{N5678}
[\hat{c}_{2}(\varrho_{\zeta_{j}}^{\perp}),\hat{c}^{\dag}_{2}(\varrho_{\xi}^{\perp})]&=&(\varrho^{\perp}_{\zeta_{j}},\varrho^{\perp}_\xi)
=(\Upsilon_{j}-\kappa\chi_{j}^{*})/(\mu\Xi^{*}_{j}).
\end{eqnarray}

Glauber's displacement operator of the form $\hat{D}(\delta;\xi)$ can be decomposed as
$\hat{D}(\delta;\xi)=\hat{D}(\delta\kappa;\varrho)\hat{D}(\delta\mu;\varrho^{\perp}_{\xi})$ enabling us to
write a coherent state of the form $|\delta;\xi\rangle$ as $|\delta;\xi\rangle =\hat{D}(\delta \kappa
;\varrho)\hat{D}(\delta \mu ;\varrho ^{\perp}_{\xi})|vac\rangle$. Moreover, from the definition of the
$\hat{P}^{c_{k}}_{m}(\varrho)$ operator, we can easily show that
$[\hat{P}^{c_{3}}_{m}(\varrho),\hat{D}(\delta \mu ;\varrho^{\perp}_{\xi})]=0$ and
$[\hat{P}^{c_{3}}_{m}(\varrho),\hat{D}^{\dagger}(\delta \mu ;\varrho^{\perp}_{\xi})]=0$. The same commutation
relation is valid for the displacement operator of mode $\varrho^{\perp}_{\zeta_{j}}$.

Using the elements of POVMs given in Eq. (\ref{Nxx12}) and the transformations
\begin{eqnarray}\label{N7896}
\hat{D}^{\dagger}(\delta\mu ;\varrho^{\perp}_{\xi})\hat{c_{3}}(\varrho^{\perp}_{\xi})\hat{D}(\delta \mu
;\varrho^{\perp}_{\xi})&=&\hat{c_{3}}(\varrho^{\perp}_{\xi})+\delta\mu\nonumber\\
\hat{D}^{\dagger}(\delta \mu
;\varrho^{\perp}_{\xi})\hat{c_{3}}^{\dagger}(\varrho^{\perp}_{\xi})\hat{D}(\delta \mu
;\varrho^{\perp}_{\xi})&=&\hat{c_{3}}^{\dagger}(\varrho^{\perp}_{\xi})+\mu^{*}\delta^{*}\nonumber\\
\hat{D}^{\dagger}(\delta
\mu;\varrho^{\perp}_{\xi})\hat{D}(\delta\mu;\varrho^{\perp}_{\xi})&=&\hat{D}^{\dagger}(\delta
\kappa;\varrho)\hat{D}(\lambda\kappa;\varrho)=\hat{{\cal I}}
\end{eqnarray}\noindent where $\hat{{\cal I}}$ is the identity
operator, together with similar expressions for $\hat{c_{2}}$, $\lambda$, and $\varrho^{\perp}_{\zeta_{j}}$,
we can obtain the following expression
\end{multicols}
\widetext
\begin{eqnarray}\label{dd1}
\langle \delta ;\xi | \Pi^{c_{3}}_{0}|
 \delta ;\xi \rangle &=&\exp[ -\eta| \alpha \kappa|^{2}/2]\nonumber \\
\langle \lambda ;\xi|\Pi^{c_{2}} _{1}|\lambda;\xi\rangle &=&\frac{1}{2}\eta|
\alpha\kappa|^{2}\exp[-\eta|\alpha\kappa|^{2}/2]\nonumber \\
\langle \delta ;\xi |\hat{c_{3}}(\zeta_{j})\Pi^{c_{3}}_{0}|\delta;\xi\rangle&=&\frac{1
}{\sqrt{2}}\alpha\left(\Upsilon_{j}-\eta\kappa \chi_{j}^{\ast }\right) \exp[-\eta| \alpha \kappa |
^{2}/2]\nonumber \\
\langle \lambda ;\xi|\hat{c_{2}}(\zeta_{j})\Pi^{c_{2}} _{1}| \lambda ;\xi
\rangle&=&\frac{i}{2\sqrt{2}}\eta\alpha
\left[~2\kappa\chi_{j}^{\ast}+|\alpha\kappa|^2(\Upsilon_{j}-\eta\kappa \chi_{j}^{\ast })~\right] \exp
[-\eta|\alpha\kappa|^{2}/2]\nonumber \\
\langle \delta ;\xi|\hat{c_{3}}(\zeta_{j})\Pi^{c_{3}}_{0}\hat{c_{3}}^{\dagger}(\zeta_{j})|\delta;\xi\rangle
&=&\frac{1}{2}\left[~2(1-\eta|\chi_{j}|^{2})+|\alpha(\Upsilon_{j}-\eta \kappa \chi_{j}^{\ast})|
^{2}~\right]\exp[-\eta|\alpha
\kappa|^{2}/2]\nonumber \\
\langle\lambda ;\xi| \hat{c_{2}}(\zeta_{j})\Pi^{c_{2}}_{1}\hat{c_{2}}^{\dagger}(\zeta_{j})|\lambda;\xi\rangle
&=&\frac{1}{4}\eta
\left[~2|\chi_{j}|^{2}(2-|\alpha\Upsilon_{j}|^2-3\eta|\alpha\kappa|^2)+|\alpha|^2\left(2|\kappa+\chi_{j}\Upsilon_{j}|^{2}+|\alpha\kappa(\Upsilon_{j}-\eta\kappa\chi_{j}^{*})|^{2}\right)\right]\nonumber\\
&&~~~~~~~~\times\exp[ -\eta| \alpha \kappa| ^{2}/2].
\end{eqnarray}
\widetext
\begin{multicols}{2}
\noindent Equation (\ref{dd1}) together with Eq. (\ref{N24}) clearly shows that the output state is dependent
on how well the modes of the input lights (single photon and coherent states) are matched to the modes of
each other and to the modes of the photon counting detectors.

Using Eq. (\ref{dd1}) in  Eqs. (\ref{N23})-(\ref{N24}), and defining the normalization parameter as
\begin{eqnarray}\label{N31bc}
{\cal{N}}_{0}&=&\sum_{j}p_{j}\left[|\chi_{j}|^{2}+|\alpha\kappa|^{2}(2-\eta|\chi_{j}|^{2})\right],
\end{eqnarray} \noindent the output density operator $\hat{\rho}_{\rm
out}$ and the probability of correct detection event ${\rm P}_{10}$ can be, respectively, written as
\begin{eqnarray}\label{N31bb}
\hat{\rho}_{\rm
out}&&={\cal{N}}^{-1}_{0}\sum_{j}p_{j}\left[~\left(|\alpha\kappa|^{2}(1-\eta|\chi_{j}|^{2})+|\chi_{j}|^{2}\right)
~| 0\rangle _{b1 b1}\langle 0|\right.\nonumber \\
&&\left.~~~~~~~~+\kappa^{\ast}\alpha ^{\ast }\chi_{j}~|0\rangle _{b1 b1}\langle 1;\zeta
_{j}|+\alpha\kappa\chi^{\ast}_{j}~ |1;\zeta
_{j}\rangle _{b1 b1}\langle 0 |\right.\nonumber \\
&&\left.~~~~~~~~+|\alpha \kappa|^{2}~|1;\zeta _{j}\rangle _{b1 b1}\langle 1;\zeta _{j}|~\right]
\end{eqnarray}\noindent and
\begin{equation}\label{N9731}
{\rm P}_{10}=\frac{1}{4}\eta{\cal{N}}_{0}\exp[-\eta|\alpha\kappa|^2].
\end{equation}
Then the fidelity of the truncation process can be found using Eqs. (\ref{N26})-(\ref{N31bb}) as
\begin{eqnarray}\label{N26bcd}
F=\frac{(1-\eta|\alpha\kappa|^2)\Gamma_{\chi}+|\alpha|^2|\alpha\kappa|^2\Gamma_{\Upsilon}+2|\alpha|^2\Gamma_{m}+|\alpha\kappa|^2}
{(1+|\alpha|^{2})\left[~(1-\eta|\alpha\kappa|^{2})\Gamma_{\chi}+2|\alpha\kappa|^{2}~\right]}
\end{eqnarray}\noindent where we have used
\begin{eqnarray}\label{N2347}
\Gamma_{\chi}&=&\sum_{j}p_{j}|\chi_{j}|^{2}, ~~~~~~\Gamma_{\Upsilon}=\sum_{j}p_{j}|\Upsilon_{j}|^{2}\nonumber\\
\Gamma_{m}&=&\sum_{j}p_{j}~{\rm Re}[\kappa^{*}\Upsilon_{j}\chi_{j}].
\end{eqnarray}

For the photon-counting detectors that are matched to $\xi$-mode, that is $\varrho=\xi$ implying $\kappa=1$
and $\chi_{j}=\Upsilon_{j}^{*}$, fidelity of the truncation process is found as
\begin{eqnarray}\label{N31be}
&&F=1-\frac{|\alpha|^{2}}{1+|\alpha|^{2}}~\frac{(1+2|\alpha|^{2})-|\gamma
_{0}|^{2}(1+|\alpha|^{2}(1+\eta))}{2|\alpha |^{2}+|\gamma_{0}|^{2}\left[1-\eta|\alpha|^{2}\right]}
\end{eqnarray}\noindent
where we have defined the mode-match parameter, with the help of Eqs. (\ref{bbb6})-(\ref{kkk2}), as
\begin{eqnarray}\label{N31bc1}
|\gamma_{0}|^{2}&=&\sum_{j}p_{j}|\langle 1;\xi|1;\zeta_{j}\rangle|^2=\sum_{j}p_{j}|\Upsilon_{j}|^2.
\end{eqnarray}\noindent   The density operator can be calculated by substituting Eqs. (\ref{N31bc1}) in Eqs. (\ref{N31bb})-(\ref{N31bc}).
For a pure one-photon state input at the $\hat{a}_{1}$ port of BS1, the density matrix simplifies into
\begin{eqnarray}\label{N31d}
\hat{\rho}_{\rm out}={\cal{N}}^{-1}_{1} \bordermatrix{& &\cr
           & |\alpha|^{2}+(1-\eta|\alpha|^{2})|\gamma_{0}|^{2} &\alpha ^{\ast }\gamma_{0}^{\ast }\cr
           & & \cr
           &\alpha \gamma_{0} &| \alpha| ^{2}\cr}
\end{eqnarray}\noindent where ${\cal{N}}_{1}=2|\alpha| ^{2}+(1-\eta|\alpha|^{2})|\gamma_{0}|^{2}$ and
$|\gamma_{0}|^{2}=|(\zeta,\xi)|^{2}$.

If the detectors can resolve only the mode of the one-photon state, ($\varrho=\zeta_{j}$), the expression for
fidelity can be obtained from Eq. (\ref{N26bcd}) by substituting $\chi_{j}=1$ and $\kappa=\Upsilon_{j}$. When
the mode-mismatch parameter ($|\gamma_{1}|^{2}=1-|\gamma_{0}|^2$), equals to one, the fidelity of truncation
becomes $1/(1+|\alpha|^2)$ independent of the detection efficiency, solely dependent on the intensity of the
input coherent light to be truncated. With increasing $|\alpha|^{2}$, fidelity of truncation decreases. On
the other hand, $P_{10}$ will take the value $\eta /4$ independent of $|\alpha|^{2}$. For the special case of
$\eta=1$, the expression for fidelity simplifies to
\begin{equation}\label{N35}
  F=1-\frac{|\alpha \gamma_{1}|^2}{1+|\alpha|^{2}}.
\end{equation} \noindent where the linear dependence of fidelity on mode-mismatch parameter is clearly seen.

The density matrix of the truncated output state, when the input one-photon state is a pure one, becomes
\begin{eqnarray}\label{N33}
 \hat{\rho}_{\rm out}={\cal{N}}^{-1}_{2}\bordermatrix{& &\cr
           &1+(1-\eta)|\alpha\gamma _{0}|^{2}& \alpha^{*} \gamma_{0}^{*} \cr
           & & \cr
           &  \alpha \gamma_{0}  & |\alpha \gamma_{0}|^{2} \cr}
\end{eqnarray}
\noindent with ${\cal{N}}_{2}=1+(2-\eta)|\alpha\gamma _{0}|^{2}$. The density matrix given in Eq. (\ref{N33})
shows that mode-match parameter affects both the diagonal and off-diagonal terms of the density matrix as
well as the probability of proper detection $P_{10}$. From Eq. (\ref{N31bb}), it is clearly seen that when
$\varrho$ is set to $\zeta$, the amplitude of the input coherent state $|\alpha;\xi\rangle$ is re-scaled with
the amount of overlap between $\xi$ and $\zeta$ modes. This corresponds to the case where an input of the
form $|\alpha \gamma_{0};\zeta\rangle$ is used as the input coherent state.

If the detectors cannot resolve the mode $\zeta_{j}$ of the one-photon state, that is
$\zeta_{j}=\varrho^{\perp}_{\zeta_{j}}$ hence $\chi_{j}=0$, the output state becomes independent of the
photon counting process which results in a classical mixture of vacuum and one-photon states. On the other
hand, when $\xi=\varrho^{\perp}_{\xi}$ with $\kappa=0$, the quantum state at the output will be vacuum
because the detected photon will always originate from the input single photon state.

\subsubsection{Mode-unresolving detectors}
Since the detectors cannot resolve the mode, photons of any mode are registered by the detectors. In a
one-photon detection event, the registered photon could have been in either of the modes. A no-photon
detection event would imply that detector has not registered any photon of neither of the modes. Then the
elements of the POVM can be written as in Eq. (\ref{Nxx12}) with $\hat{P}^{c_{k}}_{m}(\varrho)$ replaced by
$\hat{P}^{c_{k}}_{m}$ resulting in
\begin{eqnarray}
{\rm \Pi}_{0}^{c_{3}}&=&\sum_{m=0}^\infty (1-\eta)^{m}\hat{P}^{c_{3}}_{m}\nonumber \\
{\rm \Pi}_{1}^{c_{2}}&=&\sum_{m=0}^\infty m \eta(1-\eta)^{m-1}\hat{P}^{c_{2}}_{m}\label{N38a}
\end{eqnarray}\noindent where $\hat{P}^{c_{k}}_{m}$ is the projection operator to the subspace with $m$ photons in total.
Consequently, the output density operator and the probability of correct detection event $P_{01}$ become
\begin{eqnarray}\label{N31b}
\hat{\rho}_{\rm out}&&={\cal{N}}^{-1}_{3}\sum_{j}p_{j}\left[~[1+\left| \alpha \right| ^{2}\left( 1-\eta \right)]~| 0\rangle _{b1 b1}\langle 0|\right.\nonumber \\
&&\left.~~~~+\alpha ^{\ast } \Upsilon _{j}^{\ast }~|0\rangle _{b1 b1}\langle 1;\zeta _{j}|+\alpha \Upsilon
_{j}~ |1;\zeta
_{j}\rangle _{b1 b1}\langle 0 |\right.\nonumber \\
&&\left.~~~~+\left| \alpha \right| ^{2}|1;\zeta _{j}\rangle _{b1 b1}\langle 1;\zeta _{j}|~\right]
\end{eqnarray}\noindent and
\begin{equation}\label{Ngh567}
P_{10}=\frac{1}{4}\eta{\cal{N}}_{3}\exp(-\eta|\alpha|^2)
\end{equation} \noindent with ${\cal{N}}_{3}=1+\left| \alpha \right| ^{2}\left( 2-\eta \right)$.  Then the fidelity of
truncation can be calculated from Eqs. (\ref{N26}) and (\ref{N31b}) as
\begin{eqnarray}\label{N31d}
  F=1-\frac{|\alpha|^{2}[2+|\alpha|^{2}(2-\eta)-|\gamma _{0}|^{2}
 (2+|\alpha|^{2})]}{(1+|\alpha|^{2})[1+|\alpha|^{2}(2-\eta)]}
\end{eqnarray}
where we have used $|\gamma_{0}|^{2}=\sum_{j}p_{j}|\Upsilon_{j}|^{2}$ from which the amount of mode-mismatch
is calculated as $|\gamma_{1}|^{2}=1-|\gamma_{0}|^{2}$. For this type of detectors, it is observed that (i)
mode mismatch between the single photon input and the coherent input affects only the off-diagonal elements
of the output density matrix, (ii) fidelity of the truncation process decreases linearly with increasing
$|\gamma_{1}|^{2}$, (iii) the rate of decrease in the fidelity with respect to $|\gamma_{1}|^{2}$ is higher
for higher values of $\eta$ at a constant $|\alpha|^{2}$, and (iv) for $|\gamma_{1}|^{2}\leq
(1+|\alpha|^{2})/(2+|\alpha|^{2})$, increasing $\eta$ can partially compensate the mode-mismatch effect on
the fidelity and increase the value of fidelity, however for higher $|\gamma_{1}|^{2}$ values, increasing
$\eta$ causes slight decrease in fidelity.

\subsection{Conventional photodetectors }

Conventional photodetectors (CPs) that are available in the market cannot perform the ideal measurement of
photon number counting. The avalanche process taking place in the photodetectors makes it difficult to
discriminate between the presence of single photon and more photons. The outcomes of such a detector can be
either ``YES", when any number of photons are incident on the photodetector and cause a ``click" or ``NO"
when no photons are detected. Moreover, CPs cannot resolve the mode of the incoming photon and thus show a
``click" for photons belonging to any mode. Then for the QSD scheme, where there are lights with different
mode profiles incident on the detectors,
\begin{figure*}[h]
\hspace{18mm} (a) \hspace{3.7cm}(b)

\vspace{-1mm} \hspace*{-5mm} \epsfxsize=4.6cm \epsfbox{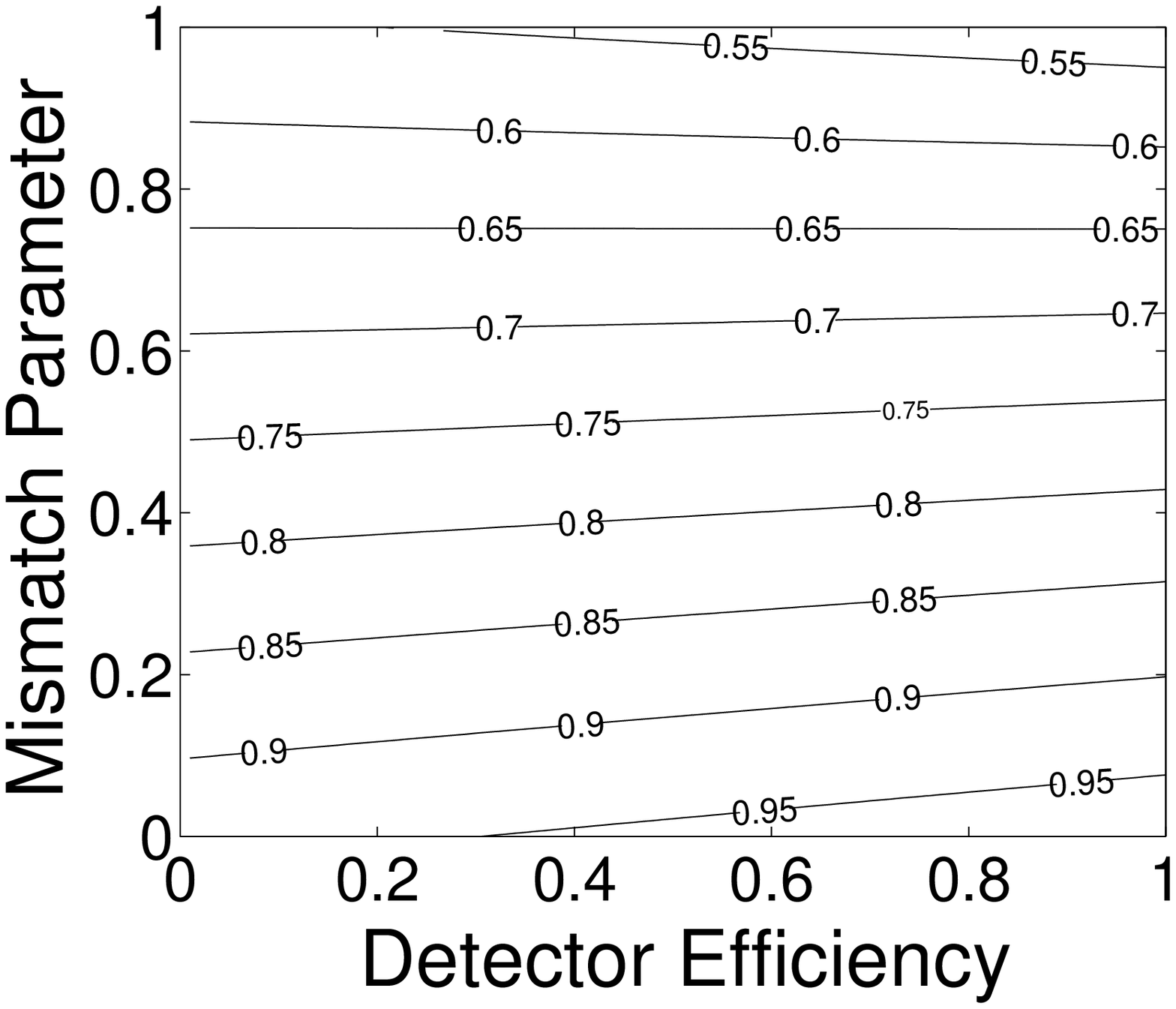}\hspace*{-1mm} \epsfxsize=4.3cm
\epsfbox{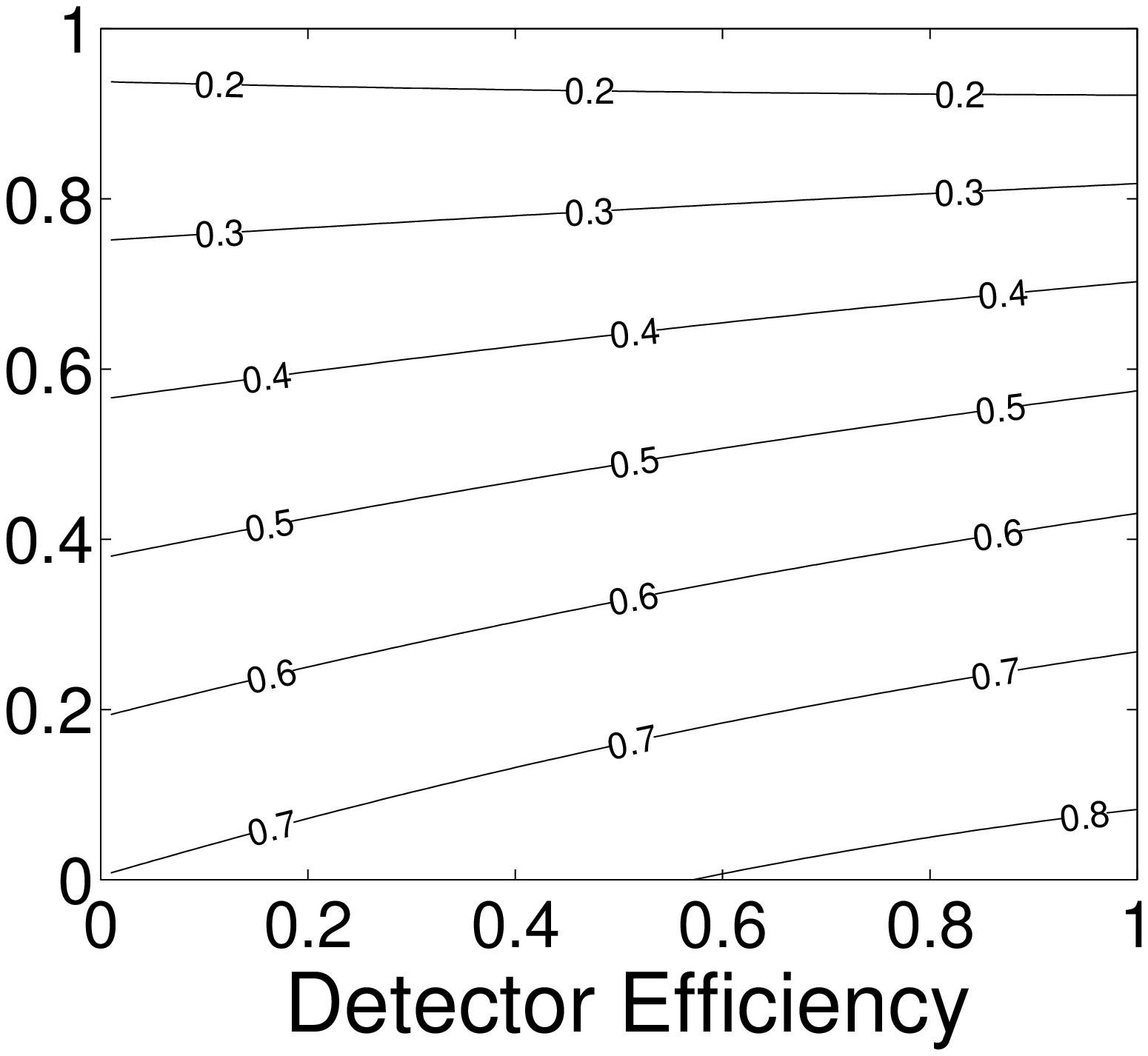}\vspace{8mm} \caption{Effect of mode-mismatch parameter $|\gamma_{1}|^{2}$ and the
detector efficiency $\eta$ on the fidelity of state truncation using conventional photon detectors. The
intensity of input coherent light to be truncated are  (a) $|\alpha|^{2}=0.4$, and (b) $|\alpha|^{2}=2.5$.
\label{Fig2}}
\end{figure*}\noindent the measurement can be described by $\Pi_{0}^{c3}$ which is the same as given in
(\ref{N38a}), and by $\Pi_{1}^{c2}$ that can be written as
\begin{eqnarray}
\Pi _{1}^{c2}&=&1-\Pi_{0}^{c2}\nonumber\\
&=&1-\sum_{m=0}^\infty (1-\eta)^{m}\hat{P}^{c_{2}}_{m}.\label{N39}
\end{eqnarray} \noindent Consequently, the elements of the density operator for the generated output state,
and $P_{10}$ are found as
\begin{eqnarray}\label{N41}
\hat{\rho}_{\rm out}&&=\sum_{j}p_{j}\left[(2(2-\eta)+
|\eta \alpha \Upsilon _{j}|^{2}-4x(1-\eta))| 0\rangle _{b1 b1}\langle 0|\right.\nonumber \\
&&\left.~~~~+2\eta\alpha ^{\ast }\Upsilon _{j}^{\ast} ~|0\rangle _{b1 b1}\langle 1;\zeta _{j}|+2\eta
\alpha\Upsilon _{j}~ |1;\zeta
_{j}\rangle _{b1 b1}\langle 0 |\right.\nonumber \\
&&\left.~~~~+4(1-x)~|1;\zeta _{j}\rangle _{b1 b1}\langle 1;\zeta _{j}|~\right]\times{\cal{N}}^{-1}_{4},
\end{eqnarray}\noindent and
\begin{equation}\label{mn279}
P_{10}=\frac{x}{8}{\cal{N}}_{4}
\end{equation} \noindent where ${\cal{N}}_{4}=\sum_{j}p_{j}\left[~2[4(1-x)+\eta(2x-1)]+|\eta \alpha \Upsilon
_{j}|^{2}~\right]$ and $x=\exp(-\eta|\alpha| ^{2}/2)$.  Then the fidelity of the truncation process can be
calculated using (\ref{N26}) as
\begin{eqnarray}\label{N42}
F&=&1-\frac{1}{(1+|\alpha|^{2}){\cal{N}}_{4}}\left \{~4\left(1-x\right)\right. \nonumber\\
&&~~~~~~~~~\left. +|\alpha|^2\left[2(4-\eta)-4x(2-\eta) \right .\right. \nonumber \\
&&~~~~~~~~~~~~~\left.\left. + |\gamma_{0}|^{2}\left(|\alpha|^{2}\eta^{2}-4(1-x+\eta)\right)\right]\right\}.
\end{eqnarray}

If the intensity of the coherent light to be truncated is $|\alpha|^2=1$, fidelity of the process for
$\eta=1$ drops from $0.94$ to $0.28$ when $|\gamma_{1}|^2$ increases from zero (perfect match between the
lights) to one (complete mismatch). In the same way, probability of correct detection events drops from $\sim
0.35$ to $\sim 0.27$. When the density matrix is analyzed for this condition, it is seen that for the
complete mismatch case, the off-diagonal elements become zero and the output state is a classical mixture of
vacuum and one-photon states, $0.56|0\rangle\langle 0|+0.44|1\rangle\langle 1|$.

In Fig. \ref{Fig2}, we have depicted constant-fidelity contours as a function of $|\gamma_{1}|^{2}$ and
$\eta$ for state truncation using conventional photon counters. It is seen that for low intensity input
coherent light, the effect of the mismatch on the fidelity of the truncation process is more profound than
that of the detector efficiency $\eta$.  Effect of $\eta$ on the value of fidelity and the allowable range of
mode mismatch is more significant for higher values of $|\alpha|^{2}$ than the smaller values. The amount of
mode-mismatch that can be tolerated to achieve a predetermined constant fidelity $F$, is much higher for low
intensity input coherent light than that of the high intensity coherent light.

\section{Mode-Mismatch Effects on State Preparation by QSD}

Quantum-scissors device which exploits projection synthesis can be used not only for state truncation but
preparation of arbitrary superposition of vacuum and one photon states, ${\it N_{\kappa}
}[\kappa_{0}|0\rangle+\kappa_{1}|1\rangle]$, as well, where ${\it
N_{\kappa}}=1/\sqrt{|\kappa_{0}|^{2}+|\kappa_{1}|^{2}}$. This can be achieved with high fidelity and nonzero
probability by properly choosing the intensity of the input coherent light. State preparation using QSD
scheme differs from the state truncation with the condition that in state truncation quantum state of the
input coherent light is not known, however in state preparation the quantum state of the input coherent light
is optimized to prepare a known desired state.

For photon number resolving detectors for which the elements of POVM are given in Eq.(\ref{N38a}), the
highest fidelity to the desired state may not necessarily be obtained at $\alpha=\kappa_{1}/\kappa_{0}$ due
to both the nonunit detector efficiency and the mode mismatch between the input states. So, we do not fix
$\alpha$ to be $\kappa_{1}/\kappa_{0}$ but leave it as a parameter to be optimized for the highest fidelity.
Then the fidelity of this state preparation is found as
\begin{eqnarray}\label{N43a}
F=\frac{1+2|\alpha \beta||\gamma_{0}|^{2}+|\alpha \beta
\gamma_{0}|^{2}+(1-\eta)|\alpha|^{2}}{(1+|\beta|^{2})[1+(2-\eta)|\alpha|^{2}]}~,
\end{eqnarray}\noindent where we have assumed that $\arg(\alpha)=\arg(\beta)$ with
$\beta=\kappa_{1}/\kappa_{0}$. Optimum value of $|\alpha|$ which maximizes the fidelity of state preparation
for an arbitrary $\beta$ is found as
\begin{equation}\label{N43b1}
|\alpha|=\left[\frac{(|\beta \gamma_{0}|^{2}-1)^{2}}{4|\beta|^{2}
|\gamma_{0}|^{4}(2-\eta)^{2}}+\frac{1}{(2-\eta)}\right]^{1/2}+\frac{|\beta \gamma_{0}|^{2}-1}{2|\beta|
|\gamma_{0}|^{2}(2-\eta)}.
\end{equation} \noindent It is seen that the increase in $\eta$ shifts the optimized value of $|\alpha|$ to
higher values. As the mode mismatch increases, the optimized value of $|\alpha|$ decreases, and in the
limiting case $|\gamma_{0}|^{2}=0$ ($|\gamma_{1}|^{2}=1$), it becomes zero independent of $\eta$ and
$|\beta|$.
\begin{figure*}[h]
\vspace*{-13mm}\hspace*{-10mm} \epsfxsize=9.5cm \epsfbox{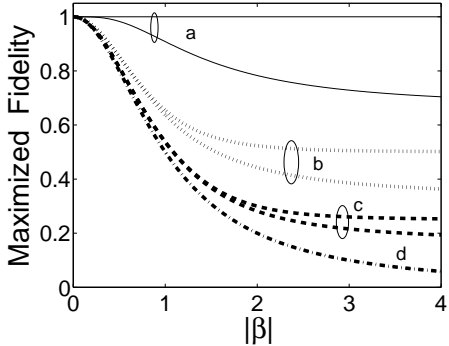}\vspace{-5mm} \caption{Effect of
mode-mismatch parameter on the maximum fidelity of state preparation with detectors of $\eta=0.5$ and
$\eta=1.0$ when $|\gamma_{1}|^{2}$ equals: a, $0$ (perfectly matched modes), b, $0.50$, c, $0.75$ and d,
$1.0$ (complete mismatch). Fidelity obtained for $\eta=1.0$ is higher than the fidelity for $\eta=0.5$.
\label{Fig3}}
\end{figure*}\noindent
Figures \ref{Fig3} and \ref{Fig4} show the results of this study for which it is understood that the relative
weight of vacuum and one-photon state in the superposition is crucial for the amount of mode mismatch that
can be tolerated for state preparation with high fidelity. If $|\beta|\leq 0.4$, fidelity values of $F\geq
0.9$ for any amount of mode mismatch with $\eta=0.5$. In this range of $|\beta|$, increasing $\eta$ does not
cause a significant improvement on the value of fidelity. The effect of $\eta$ and $|\gamma_{1}|^{2}$ is more
profound when the weight of one-photon component is dominant in the superposition, that is when
$|\beta|>1.0$. To illustrate this we consider the preparation of states with $|\beta|$ equals $0.2$ and $5$.
With $\eta=0.5$ and $|\gamma_{1}|^{2}=0.5$, the corresponding maximized fidelities are found as $0.97$ and
$0.35$. When $\eta$ is increased to $1.0$, the fidelities will be $0.97$ (no change) and $0.50$,
respectively. On the other hand, for $\eta=0.5$, increasing the mismatch parameter $|\gamma_{1}|^{2}$ from
$0.5$ to $0.8$ will cause a decrease in the fidelities which will become $0.96$ and $0.15$, respectively.

For the case of conventional photon counters as the detectors, the expression of the optimized $|\alpha|$ and
$F$ are too lengthy and complicated to give here. To have an idea of the state preparation with such
detectors, here we give some numerical values: To prepare a state with $|\beta|=0.4$, the optimum value for
$|\alpha|$ and the corresponding maximized fidelity are $0.39$ and $0.99$ when modes are completely matched
and $\eta=0.7$. However, with increasing mode mismatch amount to $|\gamma_{1}|^2=0.25$ and $0.5$, the
optimized $|\alpha|$ values decrease to $0.3$ and $0.21$ with the corresponding fidelity values of $0.94$ and
$0.90$, respectively. Further increase in the amount of mode mismatch forces the optimized $|\alpha|$ to
become much closer to zero and the fidelity to the minimum value of $0.862$. For $\eta=0.5$, the desired
state can still be prepared with a fidelity $\geq 0.90$ if the mode mismatch is kept below $0.5$ for which
the optimized $|\alpha|$ will lie in the range $0.21\leq|\alpha|\leq0.38$. Then it can be said that
superposition states, for which the
\begin{figure*}[h]
\hspace*{0mm} \epsfxsize=6cm \epsfbox{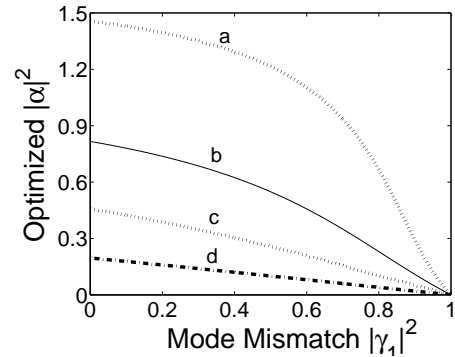}\vspace{3mm} \caption{Effect of mode mismatch parameter on
the optimized intensity of the input coherent light $|\alpha|^{2}$ at which fidelity of state preparation is
maximized for $|\beta|$ equals to: a, 0.2, b, 0.5, c, 1.0, d, 2. A detection efficiency of $\eta=0.5$ is
assumed. \label{Fig4}}
\end{figure*}\noindent vacuum component of the superposition is dominant, can be prepared with high fidelity even
with mode mismatch as much as $50\%$. On the other hand, when the one-photon component of the state becomes
dominant, the mode mismatch show a much higher deteriorating effect on the fidelity of state preparation.
When $|\beta|=2.5$ (the weights of vacuum and one-photon states are interchanged), an increase of mode
mismatch from zero to $0.25$ causes the optimized value of $|\alpha|$ to change from $1.68$ to $1.97$. This
in turn, causes the maximized fidelity to decrease from $0.71$ to $0.58$ which is a $18.3\%$ decrease much
higher than the decrease of $4.26\%$ of the case $|\beta|=0.4$.\vspace{-2mm}

\section{Mode Structures of fields and practical Considerations}

At the input of the BS2, one of the interfering fields is the input coherent light and the other field is
either a single-photon wave packet or vacuum. In case of vacuum, the mode match is not a problem. However,
when the field is single-photon then its mode must be matched as much as possible to the mode of the input
coherent light. It is seen in Eqs.(\ref{N41})-(\ref{N42}) that the fidelity of truncation process using
conventional photodetectors is a function of detector efficiency $\eta$, intensity $|\alpha|^{2}$ of the
coherent light to be truncated and the overlap of the modes of one-photon state and the coherent state which
is defined as $|\gamma_{0}|^{2}=\sum_{j}p_{j}|\Upsilon_{j}|^{2}$ when one-photon is in a mixed state of the
form given in Eq. (\ref{bbb}). Then in a practical scheme, where $\eta$ is limited with the CPs being used
and $|\alpha|^{2}$ can be set freely, an information on the value of $|\gamma_{0}|^{2}$ will enable the
calculation of fidelity to evaluate the efficiency and the quality of the truncation process. Once the mode
profiles of the input lights are characterized correctly, their overlap can be found easily using Eq.
(\ref{kkk2}) and consequently, fidelity of the process can be calculated using Eq. (\ref{N42}). Therefore, to
have an idea on the bounds of mode matching and the physical phenomena affecting the process, the
characterization of the mode structures of the coherent state $|\alpha;\xi\rangle$ and the single-photon
state $|1;\zeta\rangle$ is crucial.

In practice, the single-photon state input to the QSD scheme is prepared by a conditional measurement on a
biphoton state generated by pulsed spontaneous parametric down conversion (SPDC). In this process, a pump
beam converts spontaneously, with a small probability $\cal{O}$$(10^{-4})$, into two photons with lower
energy due to an interaction in a nonlinear crystal. The two photons, which are created almost simultaneously
within a time window that is given by the inverse of the emitted bandwidth, constitute a highly entangled
quantum state and are separated into two emission channels which are named as {\it idler} and {\it signal}.
Starting from the interaction Hamiltonian of SPDC, the biphoton state generated by a pulsed light can be
written as \cite{Ou1,Ou2,Grice,Lvovsky}
\begin{eqnarray}\label{N45}
|\varphi\rangle &=&|0\rangle_{i}|0\rangle_{s}-\I\chi\int d^{3}k_{s}d^{3}k_{i}d\omega_{s}d\omega_{i}|1;k_{i},
\omega_{i}\rangle_{i}|1;k_{s},\omega_{s}\rangle_{s} \nonumber \\
&&\times\int d^{3}k_{p}d\omega_{p} E_{p}^{(+)}(\textbf{k}_{p},\omega_{p})\int_{t_{0}}^{\infty}
dt~ \E^{\I(-\omega_{p}+\omega_{i}+\omega_{s})t} \nonumber \\
&&\times\int d^{3}r \textbf{K}(\textbf{r}) \E^{\I(\textbf{k}_{p}-\textbf{k}_{s}-\textbf{k}_{i}\textbf{r})}
+h.c.,
\end{eqnarray} \noindent where $E_{p}^{(+)}(\textbf{k}_{p},\omega_{p})$ is the positive-frequency electric field operator of the pump,  $\chi$
is proportional to the second order nonlinear susceptibility and $\textbf{K}(\textbf{r})$ describes the
volume of the nonlinear crystal with a value of one inside the crystal and zero outside. The last integral
corresponds to the Fourier transform of $\textbf{K}(\textbf{r})$ which can be written as $\textbf{K}(\Delta
\textbf{k})$ with $\Delta \textbf{k}=\textbf{k}_{p}-\textbf{k}_{s}-\textbf{k}_{i}$ and has the form of
sinc-function.  To simplify the calculations the following assumptions are made: (i) limit of the time
integration in Eq. (\ref {N45}) can be taken from $-\infty$ to $+\infty$ because we are interested in the
fields far from the crystal, thus integral becomes an impulse function $\delta
(\omega_{p}-\omega_{s}-\omega_{i})$ expressing the energy conservation in the process, (ii) crystal volume is
much larger than the spatial extent of the pump pulse inside the crystal, thus sinc-functions can be
approximated by impulse function expressing perfect phase matching, (iii) we confine ourselves to a single
spatial mode thus replace k-integrals by frequency integrals, and (iv) the pump is a pure strong coherent
light. After straightforward but lengthy calculations, we end up with the following biphoton state
\begin{equation}\label{N46}
|\varphi\rangle=|0\rangle_{i}|0\rangle_{s}-\I\chi\int
E_{p}^{(+)}(\omega_{s}+\omega_{i})d\omega_{s}d\omega_{i}|1;\omega_{i} \rangle_{i}|1;\omega_{s}\rangle_{s}~,
\end{equation} \noindent

The detection of a photon in the idler channel projects the quantum state in the signal channel into a
one-photon state. The photon in the idler channel is selected by spatial and frequency filters which
determine the mode structure of detected photon, this, in turn, will affect the mode structure of the photon
in the signal channel which is conditioned on the detection in the idler channel. Here, since only a single
spatial mode is considered we focus on the effect of the characteristics of temporal filters on the process.
Filtering operator $\hat{{\cal{F}}}_{i}$ which selects the photon in the idler channel is written as
\begin{equation}\label{N44}
\hat{{\cal{F}}}_{i}=\int d\omega_{i}F(\omega_{i})|1;\omega_{i}\rangle_{i,i}\langle1;\omega_{i}|.
\end{equation}\noindent where $F(\omega_{i})$ denotes the transmission function of the filter. Then the unnormalized
state in the signal channel becomes ${{\rm Tr}_{i}}(\hat{{\cal{F}}}_{i}|\varphi\rangle\langle\varphi|)$ where
trace is taken over idler states. Consequently, the field in the signal channel can be found as
\begin{eqnarray}\label{N47}
\hat{\rho}_{s}&=&{\rm Tr_{i}}(\hat{{\cal{F}}}_{i}|\varphi\rangle\langle\varphi|)\nonumber \\
&=& |\chi|^{2}\int d\omega_{s}d\omega^{'}_{s}d\omega_{i}F(\omega_{i})\nonumber \\
&&~~~~E_{p}^{(-)}(\omega_{s}+\omega_{i})E_{p}^{(+)}(\omega^{'}_{s}+\omega_{i})|1;\omega_{s}\rangle_{s
s}\langle1;\omega^{'}_{s}|.
\end{eqnarray}
In a practical QSD application, this one-photon state is input to BS1 after which it interferes with the
coherent state at BS2. We assume that a collimated pump field with a Gaussian spectral distribution
\begin{equation}\label{N48}
E_{p}^{(+)}(\omega)=E_{0}\exp \left[-\frac{(\omega-\omega^{o}_{p})^{2}}{2\sigma_{p}^{2}}\right]~,
\end{equation} \noindent and a Gaussian spectral filter in the idler channel with an intensity transmission
function
\begin{equation}\label{N49}
F(\omega_{i})=F_{0}\exp \left[-\frac{(\omega-\omega^{o}_{i})^{2}}{\sigma_{i}^{2}}\right]~,
\end{equation} \noindent where $2\sqrt{2}\sigma_{p}$ and $2\sigma_{i}$ are the $1/e$-widths of the pump field and the
intensity transmission function of the interference filter in the idler channel with the central frequencies
of $\omega^{o}_{p}$ and $\omega^{o}_{i}=\omega^{o}_{p}/2$. Then using (\ref{N47})-(\ref{N49}), the state in
the signal channel is found as \cite{Lvovsky}
\begin{eqnarray}\label{N50}
\hat{\rho}_{s}&=&\Gamma_{0}\int d\omega_{s}d\omega^{'}_{s} \exp \left[-\frac{\Delta X_{1}^{2}+\Delta
X_{2}^{2}}{2(\sigma_{p}^2+\sigma_{i}^2)}-\frac{\sigma_{i}^2(\omega_{s}-\omega^{'}_{s})^{2}}
{4\sigma_{p}^2(\sigma_{p}^2+\sigma_{i}^2)} \right]
\nonumber \\
&&~~~~~~~~~~~~~~~~~ \otimes|1;\omega_{s}\rangle_{s s}\langle1;\omega^{'}_{s}|~,
\end{eqnarray}\noindent where $\Gamma_{0}$ is a constant factor, $\Delta X_{1}=\omega_{s}-\omega^{0}_{p}+\omega^{0}_{i}$ and
$\Delta X_{2}$ is the same as  $\Delta X_{1}$ with $\omega_{s}$ replaced by $\omega_{s}^{'}$. A comparison of
Eq. (\ref{N50}) with Eq. (\ref{N6789}) will reveal that the term $\Gamma_{0}$ multiplied by the exponential
corresponds the mode profile, $\zeta(\omega,\omega')$ of the state in the signal channel.

The light to be truncated in the QSD scheme is in a coherent state and it is taken from the same pulsed laser
before it is frequency doubled to obtain the pump pulse of $\omega^{0}_{p}$ for SPDC. Then the coherent state
to be truncated has a spectrum with a central frequency $\omega_{c}=\omega^{0}_{p}/2$. Mode profile of the
coherent light can be found by taking its correlation function which will yield
\begin{equation}\label{N51}
\xi(\omega,\omega^{'})=\Gamma_{1}\exp
\left[-\frac{(\omega-\omega_{c})^{2}+(\omega^{'}-\omega_{c})^{2}}{2\sigma_{c}^{2}}\right]~.
\end{equation}
Assuming that the beam splitters and the propagation of the fields until they mix at BS2 do not change the
mode profile of the fields, the overlap of the one-photon state and the coherent state can be found using
\begin{eqnarray}\label{N51ab}
|\gamma_{0}|^{2}=\frac{\int d\omega d\omega^{'}\xi(\omega,\omega^{'})\zeta(\omega,\omega^{'})}{\int d\omega
\xi(\omega,\omega)\int d\omega\zeta(\omega,\omega)}
\end{eqnarray}\noindent as in Eq. (\ref{kkk2}).

If we assume that for each run of the experiment, the mode-profile of the photon in the signal channel is the
same and reproducible, the mode-match parameter can be expressed by the following simple expression
\begin{eqnarray}\label{N52}
|\gamma_{0}|^{2}=\frac{2\sigma_{c}\sigma_{p}}{(\sigma_{c}^{2}+\sigma_{p}^{2})}\frac{1}{\sqrt{1+(\sigma_{i}^{2})/(\sigma_{c}^{2}
+\sigma_{p}^{2})}}. \end{eqnarray} \noindent This expression gives the lower theoretical bound for the
temporal mode match for a QSD realization.

In practice, the data related with the filters are given as full-width-half-maximum (FWHM) of the intensity
transmission function (intensity versus wavelength) and we measure the FWHM of the intensity spectrum of the
light field. Therefore, we have to calculate the mode-match parameter using these experimentally accessible
data. Applying the narrow-bandwidth limit $\sigma_{i,p}\ll\omega^{o}_{i,p}$ and $\sigma_{c}\ll\omega_{c}$,
the relation between $\sigma_{\ell}$ of the functions in Eqs. (\ref{N48})-(\ref{N51})  and the experimentally
accessible bandwidths ($\lambda_{\ell,FWHM}$) with $\ell=c,i,p$ is given as $\sigma_{\ell}=\sqrt{1/\ln2}\pi c
(\Delta\lambda_{\ell,FWHM}/\lambda^{2}_{\ell})$ with $\lambda_{\ell}$ being the central wavelength of the
spectrum. As a preliminary experiment, for example, we have measured full-width at half-maximum (FWHM)
bandwidths as $\Delta\lambda_{c,FWHM}=7 {\rm nm}$ for a pulsed laser with central frequency $\lambda_{c}=790
{\rm nm}$ and $\Delta\lambda_{p,FWHM}=4 {\rm nm}$ after it is frequency doubled.   With these values, the
lower bound for mode match is found as $|\gamma_{0}|^{2}_{LB}\simeq 0.64$,$~0.72$ and $0.73$, respectively,
for interference filters of $\lambda_{i,FWHM}=10 {\rm nm}$, $4 {\rm nm}$, and $1 {\rm nm}$ used in the idler
channel. It is clearly seen that narrow-band filtering in the idler channel is crucial for high values of
mode overlap. The overlap of the spectrum of the modes can be further increased by using interference filter
between BS2 and the photon-counting detectors as in the scheme in \cite{Sahin01}. In that case, the problem
reduces to first filtering the signal and the coherent light spectrum with the same narrow-band filter and
then calculate their overlap. Then for a filter of the form (\ref{N49}) with a $1/e$-width of $2\sigma_{f}$,
theoretical upper bound for overlap is calculated as
\begin{eqnarray}\label{N53}
|\gamma_{0}|^{2}=\frac{2\sqrt{\mu_{c}\mu_{p}(1+2\mu_{c})}~\sqrt{1+2(\mu_{p}+\mu_{i})}~}
{(\mu_{c}+\mu_{p}+4\mu_{c}\mu_{p})\sqrt{1+\frac{2\mu_{i}(1+4\mu_{c})}{(\mu_{c}+\mu_{p}+4\mu_{c}\mu_{p})}}}~,
\end{eqnarray}\noindent where $\mu_{k}=(\sigma_{k}/\sigma_{f})^2$ with $k=i,c,p$. In the limiting cases:
(a)$~\sigma_{f}\rightarrow \infty$, this expression becomes equal to (\ref{N52}), (b)$~\sigma_{f}\rightarrow
0$, the $|\gamma_{0}|^{2}$ approaches one. Using the numerical values given above, if
$\sigma_{f}=\sigma_{i}=10 {\rm nm}$ is chosen, $|\gamma_{0}|^{2}\simeq0.83$ is obtained. For
$\sigma_{f}=\sigma_{i}=4 {\rm nm}$,  $|\gamma_{0}|^{2}$ becomes $\sim 0.86$. With filters with much narrower
bandwidths, $|\gamma_{0}|^{2}$ approaches unity. Substituting $|\gamma_{0}|^{2}=0.86$ in Eq. (\ref{N42}) with
$|\alpha|^{2}=1$ and $\eta=0.5$ gives a fidelity value of $F\simeq0.82$. When $|\alpha|^{2}=0.5$ is used, a
value of $F\simeq0.89$ is obtained.

Although in the above discussion, we have analyzed only the temporal mode matching, the expressions for the
spatial mode matching can be derived using the same procedure. It must also be noted that using very narrow
frequency and spatial filters will result in attenuation of the fields incident on the detectors causing a
decrease in the rate of having a correct detection. In a realistic experiment scheme, like the one we have
proposed in \cite{Sahin01}, very good spatial mode matching can be achieved by using single-spatial-mode
fibers after BS2 when the output modes are input to the photon-counting detectors
\cite{Kuzmich,Rarity2000,Rarity2001}. A spatiotemporal mode matching value of $\sim 0.66$ has been reported
in a quantum tomography of single-photon-state experiment \cite{Lvovsky}. Rarity et al. \cite{Rarity2001}
have reported an experimentally obtained visibility of $\sim 0.63$ in an experimental scheme similar to our
proposal \cite{Sahin01}. In another experiment performed to test Bell-type inequality for EPR state in a
homodyne measurement, Kuzmich et al. \cite{Kuzmich} have reported visibility values greater than $0.8$ by
using narrow-band filters with bandwidths 3.5nm and 6nm. Within the range of reported experimental values for
$|\gamma_{0}|^{2}$, we can predict a fidelity of $F\geq 0.7$ for state truncation and preparation using the
QSD scheme when $|\alpha|^{2}\leq 1.0$, i.e with $|\gamma_{0}|^2=0.66$, $\eta=0.5$ and $|\alpha|^{2}=0.5$, a
fidelity value of $\sim0.8$ is calculated. Higher values of $|\alpha|^2$ will reduce the attainable fidelity.

\section{CONCLUSION}

A major obstacle for the practical realization of state truncation and preparation using projection synthesis
and quantum-scissors device is the mode mismatch of the input lights to the device. In order to study this
problem and its effect on the quality of the process we have developed the pulse-mode projection synthesis,
characterize the mode of the interfering lights, and derived the analytical expressions for the output
density matrix and fidelity. The study includes not only the mode mismatch between the interfering lights but
that between them and the photodetectors, as well. POVMs for the analysis are derived and discussed. It has
been understood that mode mismatch destroys the off-diagonal elements of the output density matrix strongly
and in the limiting case of complete mismatch, off-diagonal elements become zero resulting in a classical
mixture at the output. When the intensity of the input coherent light is much lower than one, mode mismatch
and detector efficiency do not have significant effect on the output of the process. When the intensity
becomes higher, fidelity of the truncation process degrades rapidly with increasing mismatch. The same
behavior is shown to be valid for the preparation of arbitrary superpositions of vacuum and one-photon
states. It has been depicted that the intensity of the input coherent light can always be optimized to
maximize the fidelity of the preparation of a desired superposition state. When desired state has vacuum
component dominant, then effect of mismatch is not significant, however when one-photon state becomes
dominant fidelity is strongly affected by mismatch. In low mode-mismatch cases, increasing detection
efficiency increases the fidelity of truncation, however, when the mode mismatch becomes larger, the effect
of detector efficiency on the fidelity of the process decreases.

\section{ACKNOWLEDGMENTS}

We thank Takashi Yamamoto and Yu-xi Liu for stimulating discussions.

\end{multicols}
\end{document}